\documentclass[aps,pre,showpacs,twocolumn]{revtex4}
\usepackage{bm}
\usepackage{graphicx}
\bibstyle{approve.bib}
\usepackage{amssymb}
\usepackage{amsmath}
\usepackage{esint}
\usepackage{epstopdf}
\usepackage{color}
\usepackage{gensymb}
\usepackage{mathtools}
\usepackage{suffix}
\usepackage{fancyhdr}

\newcommand{\be}{\begin{equation}}
\newcommand{\ee}{\end{equation}}
\newcommand{\bea}{\begin{eqnarray}}
\newcommand{\eea}{\end{eqnarray}}

\newcommand{\br}{\mathbf{r}}

\newcommand{\bn}{\bar{n}_i}

\newcommand{\bq}{\mathbf{q}}

\newcommand{\e}{\varepsilon}

\newcommand{\tv}{\tilde{v}}

\newcommand{\may}{\tilde{h}_{ij}(k)}

\newcommand{\tG}{\tilde{G}}

\newcommand{\ce}{_{\rm c}}

\newcommand{\lo}{_{\rm l}}

\newcommand{\s}{_{\rm s}}

\newcommand{\bkap}{\bar{\kappa}}

\newcommand{\bs}{\bar{\sigma}}

\newcommand{\ew}{\varepsilon_{\rm s}}

\begin{document}

\title{Molecular-Level Evaluation of the Thermodynamic Properties of Charged Solutions from Submolar to Molar Ion Concentrations}

\author{Sahin Buyukdagli}
\address{Department of Physics, Bilkent University, Ankara 06800, Turkey\\
University of Lille, Unit\'{e} de Glycobiologie Structurale et Fonctionnelle (UGSF), CNRS UMR8576, F-59000 Lille, France}

\begin{abstract} 

The accurate estimation of the ionic activity coefficients is paramount for the characterization of the electrochemical reactions driving energy storage processes as well as the biological mechanisms involving ATP hydrolysis. The numerical evaluation of the ionic activity coefficients by computational methods is rendered difficult by the infinite dilution limit involved in the integral identity between the Helmholtz free energy and the equation of state of the electrolyte. By combining a recently introduced self-consistent Debye-H\"{u}ckel formalism and its analytically tractable low density limit, we develop a computationally efficient approach enabling the accurate calculation of the free energy and the related thermodynamic functions of aqueous electrolytes over a broad range of ionic concentration and electrostatic coupling strength. Via comparison with numerical simulations and experimental data, it is shown that our approach can predict the salt dependence of the Helmholtz free energy, the entropy, and the ionic activity coefficients from the electrostatically driven submolar ion concentrations to the hard-core dominated molar density regime.

\end{abstract}

\pacs{05.20.Jj,82.45.Gj,82.35.Rs}
\date{\today}
\maketitle   

\section{Introduction}

Electrochemical processes are ubiquitous in biological and artificial nanoscale systems embodying charged molecules dissolved into a background solvent~\cite{Book,Holm}. The characterization of the electrochemical reactions via the accurate estimation of ionic activities is thus critical for the comprehension of the thermodynamic mechanisms driving these complex systems that strongly deviate from ideality.  In this respect, the Debye-H\"{u}ckel (DH) formalism has been a seminal theory enabling the quantitative characterization of the thermodynamics of charged fluids. 

As a weak-coupling (WC) electrostatic theory, the DH approach treats interionic correlations perturbatively and also neglects the hard-core (HC) interactions in charge of stabilizing opposite-charge attraction~\cite{DH}. Consequently, the predictions of this formalism are limited to dilute aqueous monovalent solutions. While the parameter regime of concentrated liquids can be accessed via accurate numerical techniques such as Monte-Carlo (MC) simulations~\cite{Valleau,Svensson,NetzMC,NetzMC2,LevinMC,Forsman,Lamperski2025}, integral equation approaches~\cite{Blum,Henderson,Boda,Hoye,Roji}, and density functional theories~\cite{Hansen}, the need for analytical insight into electrostatically driven phenomena requires the development of mathematically transparent formalisms completing these efficient computational tools of significant numerical complexity.  

Along these lines, Attard~\cite{AttardRev} and Kjellander~\cite{Kj2,Kj1995} formulated effective electrostatic theories employing DH-like kernels whose range is determined by forcing the zeroth and second moment conditions on the pair distribution functions~\cite{Stir}. On the side of microscopic statistical mechanics, Khodolenko and Beyerlein introduced the first functional integral formulation of the three-dimensional Coulomb liquid~\cite{Kholodenko1,Kholodenko2}. Subsequently, this functional integral approach developed for bulk solutions has been extended to nano-confined electrolytes by Podgornik and Zeks~\cite{PodWKB}. Within the framework of this field-theoretic formalism, the electrostatic WC regime of the Coulomb liquid has been analyzed by systematic loop-expansion techniques~\cite{NetzLoop} and variational treatments~\cite{netzvar,HatloVar0,Buyuk2012}.  The opposite regime of electrostatic strong coupling (SC) has been characterized as well by virial expansion techniques~\cite{NetzSC} and hybrid approaches treating the distinct ionic species of the electrolyte according to their electrostatic coupling strength~\cite{Podgornik2011,Buyuk2020}. 

In the context of the WC/SC dichotomy formulated by Netz et al.~\cite{NetzLoop,NetzSC}, the splitting approach developed by Santangelo has been a pivotal achievement~\cite{Santangelo}. Santangelo's technique is based on the separation of the pairwise Coulomb interactions into a short- and a long-range component. The corresponding splitting enables the treatment of these interactions ranges by different approximations. In order to determine the value of the characteristic length splitting these two ranges, Hatlo and Lue introduced an ingenious variational method capable of interpolating self-consistently between the WC and SC regimes of inhomogeneous electrolytes confined to charged membranes~\cite{HatloSoft,HatloEpl,LueSoft}.

The thermodynamic stability of charged systems is set by the balance between electrostatic attraction and HC repulsion whose relative weight vary with the packing fraction of the charged fluid. As one approaches the regime of molar ion concentrations, the electrostatic forces dominating the system at dilute concentrations are taken over by HC interactions. Hence, the extrapolation of the predictions made by the aforementioned electrostatic theories into large ion densities requires the explicit inclusion of HC interactions into these frameworks. 

The incorporation of HC interactions into the field-theoretic formulation of charged liquids has been enabled by the virial expansion technique developed by Netz et al.~\cite{NetzHC1,NetzHC2}. Due to the intrinsic nature of the underlying virial approach that consists in expanding the free energy in terms of the ionic fugacities, the validity of this formalism is restricted to dilute ion concentrations. With the aim to access the regime of large ion concentrations where the HC contribution becomes comparable with electrostatics, we developed a cumulant-corrected DH (CCDH) formalism treating electrostatic and HC interactions on an equal footing~\cite{Buyuk2024}. Then, via the integration of a generalized version of the variational splitting approach by Hatlo and Lue~\cite{HatloSoft,HatloEpl,LueSoft}, we have improved the accuracy of the CCDH formalism in accounting for charge correlations. The resulting self-consistent DH (SCDH) formalism has been shown to reproduce accurately the ionic pair distributions, the osmotic coefficient, and the internal energy of aqueous monovalent electrolytes from submolar to molar ion densities~\cite{Soft2024,Buyuk2026}.

The energy $u_{\rm ex}$ and pressure $P$ of ionic solution can be directly calculated via the pair distributions $g_{ij}(r)$ or equivalently the correlation functions $H_{ij}(r)=g_{ij}(r)-1$ obtained from MC simulations or computational theories (see Eqs.~(\ref{eq2})-(\ref{eq3}) of the main text). However, the calculation of the Helmholtz free energy $f_{\rm ex}$ and the thermodynamic variables related to the former requires the evaluation of the numerically non-trivial integral
\be
\label{eq0}
\frac{\beta f_{\rm ex}}{2n_i}=\int_0^{n_i}\mathrm{d}n'_i\frac{\phi(n'_i)-1}{n'_i}
\ee
 of the osmotic coefficient $\phi=\beta P/(2n_i)$ over the ionic concentration $n_i$~\cite{NetzMC}. 
 
The non-triviality of the integral~(\ref{eq0}) stems from the presence of the ion density in the denominator of the integrand. As a result of this specific density dependence, the accurate evaluation of the integral necessitates the computation of the osmotic coefficient $\phi(n_i)$ with increasing numerical precision towards the infinite dilution limit $n_i\to0$. Due to this numerical complication, in our earlier works of Refs.~\cite{Soft2024,Buyuk2026}, we have not been able to use the SCDH formalism for the evaluation of the Helmholtz free energy and the thermodynamic functions obtained from the former. In the present work, we overcome this obstacle via the analytical treatment of the low density branch of the free energy integral~(\ref{eq0}). This novelty is the main progress of this article.

Our article is organized as follows. In Sec.~\ref{th}, we describe the electrolyte model and review the fundamental identities of the SCDH approach required for the computation of the thermodynamic functions. Then, we express these identities in a dimensionless form suitable for direct numerical implementation, and modify the integral on the r.h.s. of the free energy equality~(\ref{eq0}) via the analytical evaluation of its numerically problematic low density part with the loop-expanded CCDH approach. 

In Sec.~\ref{res}, the Helmholtz free energy, the excess entropy, and the ionic activity coefficients obtained from this hybrid formalism are compared with the MC simulation results of Refs.~\cite{Lamperski2025,NetzMC} and the experimental data of Ref.~\cite{Hamer}. Based on this comparative study, in Conclusions, we carry out a quantitative analysis of the validity limit of our formalism and discuss potential upgrades of the underlying theoretical framework.

\section{Theory}
\label{th}
\subsection{Review of the SCDH formalism}

We introduce here the electrolyte model and review the main identities of the SCDH formalism required for the calculation of the thermodynamic functions. We will skip the details of the derivations that have been throughly explained in our previous works~\cite{Soft2024, Buyuk2026}.

The bulk electrolyte of temperature $T$ is composed of $p$ ionic species. Each point-like ion of the species $i$ with valency $q_i$ and concentration $n_i$ is placed at the center of an impenetrable HC sphere with radius $a$. These HC charges interact via the bulk Coulomb potential $v\ce(r)=\ell_{\rm B}/r$ defined in terms of its inverse operator 
\be\label{eq1}
v\ce^{-1}(\br,\br')=-\frac{1}{4\pi\ell_{\rm B}}\nabla^2\delta^3(\br-\br'),
\ee
where the Bjerrum length $\ell_{\rm B}=e^2/(4\pi\e_0\ew k_{\rm B}T)$ includes the elementary charge $e$, the dielectric permittivity $\e_0$ of vacuum, the relative permittivity $\ew$ of water, and the Boltzmann constant $k_{\rm B}$. Moreover, electroneutrality implies the constraint
\be
\label{eq1II}
\sum_{i=1}^pn_iq_i=0
\ee
on the ionic concentrations.

For the corresponding Coulomb liquid, the energy density and pressure respectively read~\cite{Soft2024,Hansen}
\bea
\label{eq2}
\beta u_{\rm ex}&=&2\pi\ell_{\rm B}\sum_{i=1}^p\sum_{j=1}^pn_in_jq_iq_j\int_a^\infty\mathrm{d}rrH_{ij}(r);\\
\label{eq3}
\beta P&=&\sum_{i=1}^pn_i+\frac{2\pi a^3}{3}\left(\sum_{i=1}^pn_i\right)^2\\
&&+\frac{2\pi a^3}{3}\sum_{i=1}^p\sum_{j=1}^pn_in_jH_{ij}(a^+)+\frac{1}{3}\beta u_{\rm ex},\nonumber
\eea
where $\beta=1/(k_{\rm B}T)$ is the inverse thermal energy. Within the functional integral representation of the Grand-Canonical partition function at the basis of the SCDH approach, the total correlation function $H_{ij}(r)$ in Eqs.~(\ref{eq2})-(\ref{eq3}) is calculated via a variational splitting scheme followed by a first-order cumulant expansion of the Schwinger-Dyson identities. The splitting approach is based on the separation of the Coulomb potential into a short- and a long-range interaction component,
\be
\label{eq4}
v\ce(\br,\br')=v\s(\br,\br')+v_{\rm l}(\br,\br').
\ee
In Eq.~(\ref{eq6}), the long-range component is defined in terms of its inverse given by the specific differential form
\be\label{eq5}
v^{-1}\lo(\br,\br')=\left(1-\sigma^2\nabla^2+\sigma^4\nabla^4\right)v\ce^{-1}(\br,\br')
\ee
that has been originally introduced for the treatment of confined electrolytes in Refs.~\cite{HatloSoft,HatloEpl}. In Eq.~(\ref{eq5}), the quartic differential operator acting on the inverse Coulomb operator~(\ref{eq1}) and including the variational splitting length $\sigma$ filters out the short-range ion correlations. In the present article, the Fourier transform (FT) and the inverse FT of the general function $g(\br)$ are respectively defined in the form $\tilde{f}(q)=\int\mathrm{d}^3\br f(\br)e^{-i\bq\cdot\br}$ and $f(\br)=(2\pi)^{-3}\int\mathrm{d}^3\bq\;\tilde{f}(q)e^{i\bq\cdot\br}$. Thus, inverting Eq.~(\ref{eq5}) in Fourier space, and using the constraint~(\ref{eq4}), the FT of the short- and long-range potential components follow as
\bea
\label{eq6}
\hspace{-7mm}\tv_{\rm s}(q)&=&\tv\ce(q)\frac{\sigma^2q^2+\sigma^4q^4}{1+\sigma^2q^2+\sigma^4q^4};\hspace{3mm}\tv\ce(q)=\frac{4\pi\ell_{\rm B}}{q^2};\\
\label{eq7}
\hspace{-7mm}\tv_{\rm l}(q)&=&\tv\ce(q)\left(1+\sigma^2q^2+\sigma^4q^4\right)^{-1}.
\eea
Evaluating now the inverse FT of Eqs.~(\ref{eq6})-(\ref{eq7}), the corresponding potentials follow in real space as
\bea
\label{eq6II}
v_{\rm s}(r)&=&\frac{\ell_{\rm B}}{r}\left\{\cos\left(\frac{r}{2\sigma}\right)+\frac{1}{\sqrt{3}}\sin\left(\frac{r}{2\sigma}\right)\right\}e^{-\frac{\sqrt{3}r}{2\sigma}};\\
\label{eq7II}
v_{\rm l}(r)&=&\frac{\ell_{\rm B}}{r}-v_{\rm s}(r).
\eea

After performing the first-order cumulant expansion of the SD identities, the explicit form of the total correlation function in Eqs.~(\ref{eq2})-(\ref{eq3}) follow in the form
\be
\label{eq8}
H_{ij}(r)=h_{ij}(r)+\left[h_{ij}(r)+1\right]T_{ij} (r).
\ee
In Eq.~(\ref{eq8}), the Mayer function
\be
\label{eq9}
h_{ij}(r)=e^{-q_iq_j\left[G_{{\rm l},0}(r)+v\s(r)\right]}\theta(r-a)-1
\ee
includes the short-range potential component~(\ref{eq6II}), and the screened counterpart of the long-range potential~(\ref{eq7II}) dressed by many-body interactions,
\be
\label{eq10}
G_{{\rm l},0}(r)=\frac{\ell_{\rm B}}{r}\sum_{\gamma=\{0,\pm\}}\frac{e^{ik_ir/\sigma}}{3k_\gamma^4+2k_\gamma^2+1}.
\ee
Eq.~(\ref{eq10}) involves the auxiliary parameters
\bea
\label{eq11}
k_0&=&\frac{1}{\sqrt3}\left\{\frac{l}{2^{1/3}}-\frac{2^{4/3}}{l}-1\right\}^{1/2};\\
\label{eq12}
k_\pm&=&\pm\frac{1}{\sqrt6}\left\{-2+\frac{2^{4/3}}{l}-\frac{l}{2^{1/3}}\right.\\
&&\hspace{1.2cm}\left.\pm i\left(\frac{2^{4/3}\sqrt{3}}{l}+\frac{\sqrt{3}l}{2^{1/3}}\right)\right\}^{1/2}.\nonumber
\eea
In Eqs.~(\ref{eq11})-(\ref{eq12}), we also introduced the additional parameter $l=\left(7-27t^2+3^{3/2}\sqrt{27t^4-14t^2+3}\right)^{1/3}$ with $t=\kappa\sigma$, where the DH parameter $\kappa$ is defined by
\be\label{eq13}
\kappa^2=4\pi\ell_{\rm B}\sum_{i=1}^pn_iq_i^2.
\ee

The total correlation function~(\ref{eq8}) includes as well the auxiliary function $T_{ij}(r)$ given by the Fourier integral
\bea
\label{eq14}
T_{ij}(r)&=&\int_0^\infty\frac{\mathrm{d}qq^2}{2\pi^2}\frac{\sin(qr)}{qr}\\
&&\hspace{7mm}\times\left\{\sum_{k=1}^pn_k\left[\tilde{h}_{ik}(q)\tilde{h}_{kj}(q)-q_iq_jq_k^2\tG_{{\rm l},0}^2(q)\right]\right.\nonumber\\
&&\left.\hspace{1.3cm}-q_iq_j\tG_{{\rm l},1}(q)\right\},\nonumber
\eea
where the FT of the Mayer function~(\ref{eq9}) is
\bea
\label{eq15}
\may&\hspace{-2mm}=\hspace{-2mm}&-\frac{4\pi}{q^3}\left[\sin(qa)-qa\cos(qa)\right]\\
&&+4\pi\int_a^\infty\hspace{-2mm}\mathrm{d}rr^2\frac{\sin(qr)}{qr}\left\{e^{-q_iq_j\left[v\s(r)+G_{{\rm l},0}(r)\right]}-1\right\},\nonumber
\eea
and the FT of the many-body-dressed long-range potential~(\ref{eq10}) and its first order cumulant correction read
\bea
\label{eq15II}
\tG_{{\rm l},0}(q)&\hspace{-2mm}=\hspace{-2mm}&\left\{\tv_{\rm l}(q)+\sum_{i=1}^pn_iq_i^2\right\}^{-1};\\
\label{eq16}
\tG_{{\rm l},1}(q)&\hspace{-2mm}=\hspace{-2mm}&-\tG_{{\rm l},0}^2(q)\hspace{-1mm}\sum_{i,j}n_in_jq_iq_j\hspace{-1mm}\left\{\may+q_iq_j\tG_{{\rm l},0}(q)\right\}.\nonumber\\
\eea
Finally, the splitting length $\sigma$ satisfies the integral relation
\bea
\label{eq17}
&&\sum_{i,j}n_in_jq_iq_j\int_0^\infty\mathrm{d}qq^2\left[\may+q_iq_j\tG_{{\rm l},0}(q)\right]\nonumber\\
&&\hspace{3cm}\times\left[1-\tG_{{\rm l},0}^2(q)\tv_{\rm l}^{-2}(q)\right]\partial_\sigma\tv_{\rm l}(q)\nonumber\\
&&=0
\eea
obtained from the invariance of the grand potential under the variation of the parameter $\sigma$.

\subsection{Recasting the SCDH Identities in Dimensionless Form}
\label{adim}

In the remainder, we consider symmetric $q_i:q_i$ electrolytes and recast the characteristic identities of the SCDH formalism in dimensionless form. For the sake of reproducibility of the results, this rescaling is introduced below in the order that should be followed for the numerical implementation of the SCDH identities. 

To this aim, we introduce first the dimensionless distance $u=r/a$ and splitting length $\bar\sigma=\sigma/a$, the rescaled ion density $\bn=n_ia^3$, and the dimensionless Fourier variable $k=qa$ and FT $\bar{f}(k)=\tilde{f}(q)/a^3$. In terms of these dimensionless variables, the short-range and screened long-range potentials in Eq.~(\ref{eq6II}) and~(\ref{eq10}) read
\bea\label{eq18}
\hspace{-5mm}v\s(u)&=&\frac{\Gamma}{q_i^2u}\left\{\cos\left(\frac{u}{2\bs}\right)+\frac{1}{\sqrt{3}}\sin\left(\frac{u}{2\bs}\right)\right\}e^{-\frac{\sqrt{3}u}{2\bs}};\\
\label{eq19}
\hspace{-5mm}G_{{\rm l},0}(u)&=&\frac{\Gamma}{q_i^2u}\sum_{\gamma=\{0,\pm\}}\frac{e^{ik_iu/\bs}}{3k_\gamma^4+2k_\gamma^2+1}.
\eea
In Eqs.~(\ref{eq18})-(\ref{eq19}), $\Gamma=q_i^2\ell_{\rm B}/a$ stands for the electrostatic coupling parameter. Moreover, the auxiliary parameter $t$ involved in the dimensionless parameters $k_\gamma$ of Eqs.~(\ref{eq11})-(\ref{eq12}) is now given by $t=\bkap\bs$, where the rescaled DH parameter has been defined as 
\be\label{eq19II}
\bkap\equiv\kappa a=\sqrt{8\pi\Gamma\bn}.
\ee

The Mayer function~(\ref{eq9}) can be now expressed in terms of the potentials~(\ref{eq18})-(\ref{eq19}) as
\be\label{eq20}
h_{+\pm}(u)=e^{\mp q_i^2\left[v\s(u)+G_{{\rm l},0}(u)\right]}\theta(u-1)-1,
\ee
and its FT~(\ref{eq15}) reads in dimensionless form 
\bea
\label{eq21}
\bar{h}_{+\pm}(k)&=&-\frac{4\pi}{k^3}\left[\sin(k)-qa\cos(k)\right]\\
&&+4\pi\hspace{-1mm}\int_1^\infty\hspace{-3mm}\mathrm{d}uu^2\frac{\sin(ku)}{ku}h_{+\pm}(u).\nonumber
\eea
Moreover, the FT~(\ref{eq7}) of the bare long-range potential, and its screened counterparts~(\ref{eq15II}) and~(\ref{eq16}) are
\bea\label{eq22}
&&\hspace{-4mm}\bar{v}\lo(k)\hspace{-.5mm}=\hspace{-.5mm}\frac{4\pi\Gamma}{q_i^2k^2F(k)};\;F(k)=1+\bs^2k^2+\bs^4k^4;\\
\label{eq23}
&&\hspace{-4mm}\bar{G}_{{\rm l},0}(k)\hspace{-.5mm}=\hspace{-.5mm}\frac{4\pi\Gamma}{q_i^2}\frac{1}{k^2F(k)+\bkap^2};\\
\label{eq24}
&&\hspace{-4mm}\bar{G}_{{\rm l},1}(k)\hspace{-.5mm}=\hspace{-.5mm}-2\bn^2q_i^2\bar{G}^2_{{\rm l},0}(k)\hspace{-1mm}\left[\bar{h}_{++}(k)\hspace{-.5mm}-\hspace{-.5mm}\bar{h}_{+-}(k)
\hspace{-.5mm}+\hspace{-.5mm}2q_i^2\bar{G}_{{\rm l},0}(k)\right].\nonumber\\
\eea
With the knowledge of the functions~(\ref{eq21})-(\ref{eq24}), the dimensionless form of the variational Eq.~(\ref{eq17}) given by
\be\label{eq25}
\int_0^\infty\hspace{-2mm}\mathrm{d}kk^2\bar{G}_{{\rm l},1}(k)\left[\bar{G}^{-2}_{{\rm l},0}(k)-\bar{v}^{-2}\lo(k)\right]\partial_{\bs}\bar{v}\lo(k)=0
\ee
can be numerically solved by a dichotomy algorithm. 

The total correlation function~(\ref{eq8}) can be now written as a function of the dimensionless distance in the form
\be
\label{eq26}
H_{ij}(u)=h_{ij}(u)+\left[h_{ij}(u)+1\right]T_{ij} (u),
\ee
where the similar- and opposite-charge components of the auxiliary function~(\ref{eq14}) can be expressed in terms of the dimensionless FTs~(\ref{eq21}) and~(\ref{eq23})-(\ref{eq24}) as
\bea
\label{eq27}
T_{++}(u)&\hspace{-2mm}=\hspace{-2mm}&\int_0^\infty\hspace{-1mm}\frac{\mathrm{d}kk^2}{2\pi^2}\frac{\sin(ku)}{ku}\left\{\bn\left[\bar{h}^2_{++}(k)+\bar{h}^2_{+-}(k)\right]\right.\\
&&\left.\hspace{2.8cm}-2q_i^4\bn\bar{G}^2_{{\rm l},0}(k)-q_i^2\bar{G}_{{\rm l},1}(k)\right\};\nonumber\\
\label{eq28}
T_{+-}(u)&\hspace{-2mm}=\hspace{-2mm}&\int_0^\infty\hspace{-1mm}\frac{\mathrm{d}kk^2}{2\pi^2}\frac{\sin(ku)}{ku}\left\{2\bn\bar{h}_{++}(k)\bar{h}_{+-}(k)\right.\\
&&\left.\hspace{2.8cm}+2q_i^4\bn\bar{G}^2_{{\rm l},0}(k)+q_i^2\bar{G}_{{\rm l},1}(k)\right\}.\nonumber
\eea
Finally, within the dimensionless notation introduced herein, the excess energy density and the osmotic pressure $\phi\equiv\beta P/(2n_i)$ of the symmetric electrolyte follow from Eqs.~(\ref{eq2})-(\ref{eq3}) in terms of the total correlation function~(\ref{eq26}) as
\bea
\label{eq29}
\hspace{-9mm}&&\frac{\beta u_{\rm ex}}{2n_i}=2\pi\bn\Gamma\int_1^\infty\mathrm{d}uu\left[H_{++}(u)-H_{+-}(u)\right];\\
\label{eq30}
\hspace{-9mm}&&\phi=1+\frac{2\pi}{3}\bn\left\{H_{++}(1^+)+H_{+-}(1^+)+2\right\}+\frac{\beta u_{\rm ex}}{6\bn}.
\eea
As expected, the thermodynamic functions~(\ref{eq29})-(\ref{eq30}) depend solely on the electrostatic coupling parameter $\Gamma$ and the dimensionless ion density $\bn$.

\subsection{Evaluation of the Helmholtz Free Energy}
\label{hel}

The calculation of the entropy and the activity coefficients requires the knowledge of the Helmholtz free energy $F$. The latter can be obtained from the definition of pressure, i.e. $P=-\partial_V F=2n_ik_{\rm B}T-\partial_V F_{\rm ex}$, where $F_{\rm ex}$ stands for the excess part of the Helmholtz free energy. Using now the definition the osmotic pressure  $\phi\equiv\beta P/(2n_i)$, the former identity yields
\be
\label{eq31}
\phi-1=-\frac{\beta}{2n_i}\frac{\partial F_{\rm ex}}{\partial V}=n_i\frac{\partial\left[\beta F_{\rm ex}/(2N_i)\right]}{\partial n_i},
\ee
where we used  the relation $d/dV=-(n_i^2/N_i)d/dn_i$ that follows from the definition of the average ion density $n_i=N_i/V$ at fixed ion number $N_i$. Integrating now Eq.~(\ref{eq31}), introducing the free energy density $f_{\rm ex}=F_{\rm ex}/V$, and passing to the dimensionless variables of Sec.~\ref{adim}, the Helmholtz free energy per particle can be finally expressed in terms of the osmotic pressure~(\ref{eq30}) as~\cite{NetzMC}
\be
\label{eq32}
\frac{\beta f_{\rm ex}}{2n_i}=\int_0^{\bn}\mathrm{d}\bn'\frac{\phi\left(\bn';\Gamma\right)-1}{\bn'}.
\ee

Owing to the presence of the ion density in the denominator of the integrand in Eq.~(\ref{eq32}), the accurate evaluation of the free energy requires the calculation of the osmotic coefficient with increasingly higher precision towards the lower bound of the integral. Due to this numerical complication occurring close to the infinite dilution limit, in our earlier works, we have not been able to use the SCDH formalism to evaluate the free energy, and the related thermodynamic functions such as the entropy, and the ionic activity coefficients whose evaluation has been limited to the CCDH approach~\cite{Buyuk2024,Buyuk2026}. 

In the present work, this numerical obstacle will be overcome via the analytical evaluation of the low density branch of the integral in Eq.~(\ref{eq32}) with the CCDH-level osmotic coefficient obtained by loop-expansion, i.e.
\be
\label{eq33}
\frac{\beta f_{\rm ex}}{2n_i}\approx\left.\frac{\beta f^{(0)}_{\rm ex}}{2n_i}\right|_{\bn^*}+\int_{\bn^*}^{\bn}\mathrm{d}\bn'\frac{\phi\left(\bn';\Gamma\right)-1}{\bn'},
\ee
where the loop-expanded form of the Helmholtz free energy derived in Appendix~\ref{loop} reads
\bea
\label{eq34}
\frac{\beta f^{(0)}_{\rm ex}}{2n_i}&=&\frac{\bkap^2}{6\Gamma}+\frac{5\bkap^4}{576\Gamma^2}\\
&&+\frac{1}{2304}\left\{96\left[\mathrm{Ei}(-2\bkap)-\mathrm{Ei}(-\bkap)-\ln(2)\right]\right.\nonumber\\
&&\hspace{1.3cm}-16+16\left(5\bkap^2+\bkap+7\right)e^{-\bkap}\nonumber\\
&&\hspace{1.3cm}-\left(20\bkap^3-18\bkap^2+6\bkap+99\right)e^{-2\bkap}\nonumber\\
&&\hspace{1.3cm}\left.+3(4\bkap+1)e^{-4\bkap}\right\}\nonumber\\
&&+\frac{\Gamma}{6912}\left\{-2125-144(2\bkap^2+2\bkap-27)e^{-\bkap}\right.\nonumber\\
&&\hspace{1.3cm}+72(2\bkap^3-3\bkap^2-22\bkap-27)e^{-2\bkap}\nonumber\\
&&\hspace{1.3cm}+16(12\bkap+13)e^{-3\bkap}\nonumber\\
&&\hspace{1.3cm}\left.+9(16\bkap^2-12\bkap-3)e^{-4\bkap}\right\},\nonumber
\eea
and the value of the cut-off density $\bn^*$ will be specified below. In Eq.~(\ref{eq34}), ${\rm Ei}(x)$ stands for the exponential integral function~\cite{math}. 

The approximation introduced in Eq.~(\ref{eq33}) is based on the fact that the low density limit of the SCDH formalism can be accurately reproduced by the CCDH approach~\cite{Buyuk2024}. This point is explicitly illustrated by the osmotic coefficient curves reported in Fig.~\ref{fig5} of Appendix~\ref{comp}. Therein, we show that the CCDH and SCDH results  coincide in the regime of low volume fractions $\eta\lesssim10^{-3}$, where $\eta=\pi n_ia^3/3$ stands for the ionic packing fraction. Hence, in the present work, the cut-off density in Eq.~(\ref{eq33}) will be set to $\bn^*=(3/\pi)\;10^{-3}$.

\begin{figure}
\includegraphics[width=1\linewidth]{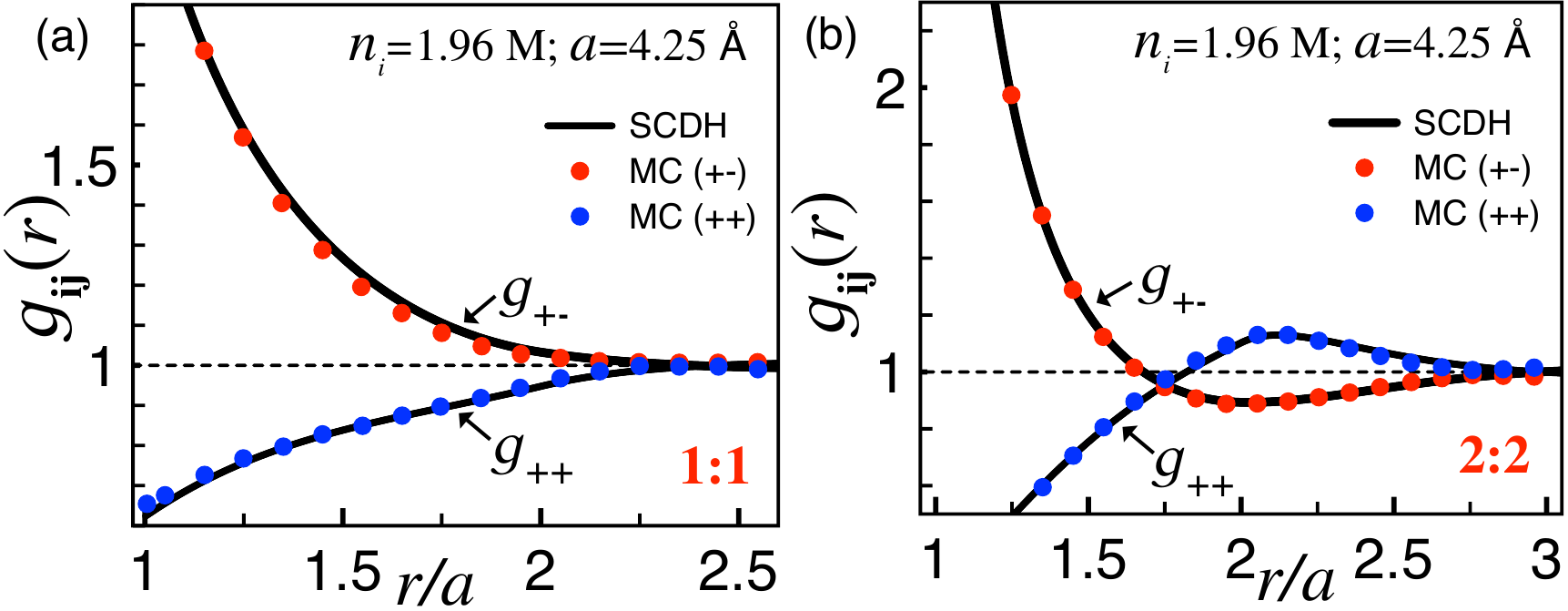}
\caption{(Color online)  Oppositely charged (+-) and similarly charged (++) pair distribution functions of (a) monovalent and (b) divalent electrolytes at concentration $n_i=1.96$ M, HC size $a=4.25$ {\AA}, temperature $T=298.15$ K, and dielectric constant $\e\s=78.5$. Solid curves: SCDH prediction $g_{ij}(r)=H_{ij}(r)+1$ from Eq.~(\ref{eq26}). Symbols: MC simulation data from Figure 1 of Ref.~\cite{Lamperski2025}.}
\label{fig1}
\end{figure}

\section{Comparison with Simulations and Experiments}
\label{res}

In this section, we compare the predictions of the SCDH formalism for the free energy~(\ref{eq33}) and the related thermodynamic functions with the MC simulation results of Refs.~\cite{Lamperski2025,NetzMC} and the experimental data of Ref.~\cite{Hamer}.

\subsection{Pair distributions}

In our previous article of Ref.~\cite{Buyuk2026}, the SCDH predictions for monovalent pair distributions have been compared with the MC simulation results of Ref.~\cite{Forsman} over a broad concentration range. Here, we compare the radial distributions obtained from the SCDH formalism with the separate MC data of Ref.~\cite{Lamperski2025} for monovalent ions, as well as for divalent ions that has not been considered in our earlier work~\cite{Buyuk2026}.

\begin{figure}
\includegraphics[width=1\linewidth]{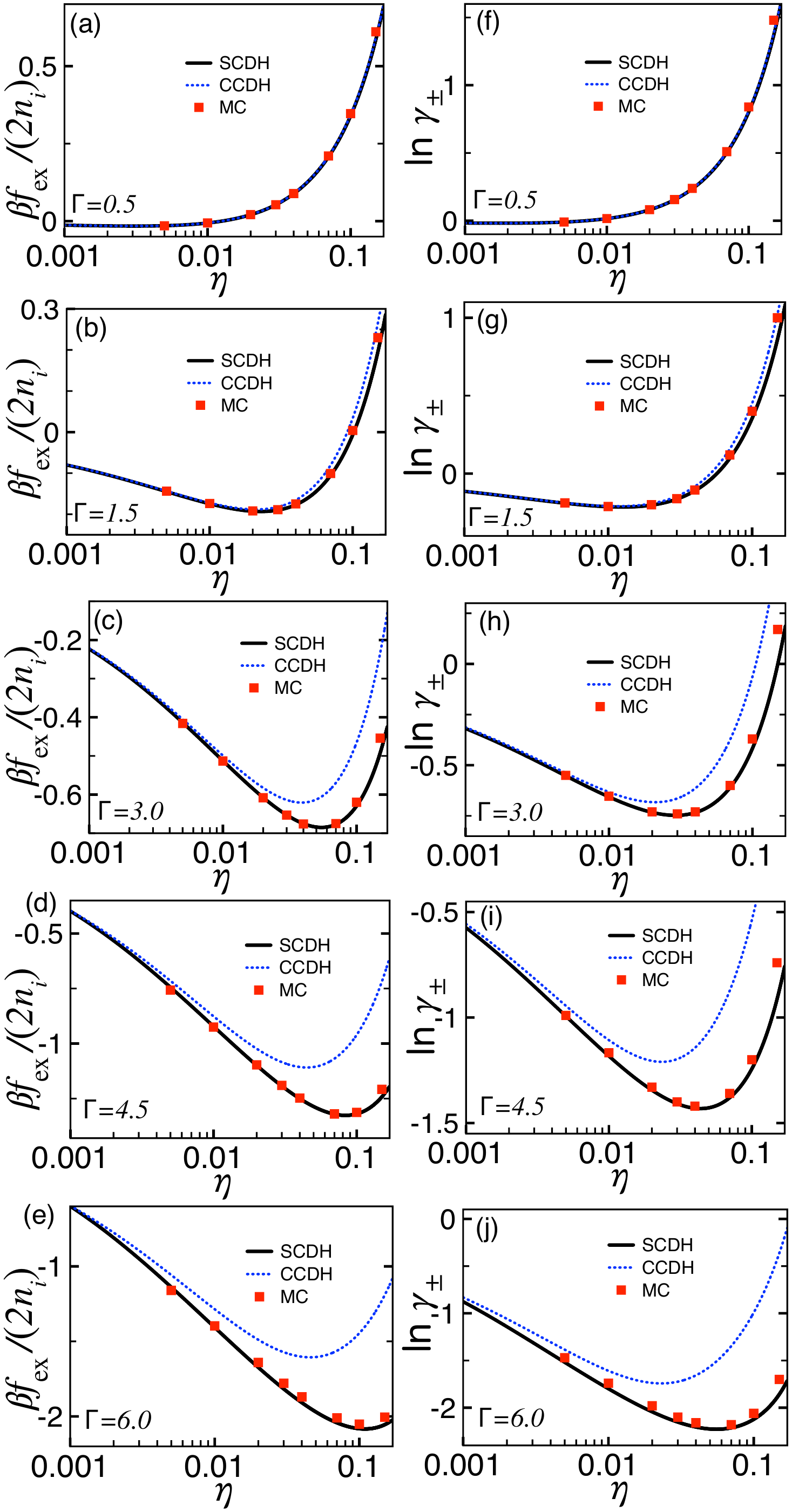}
\caption{(Color online)  Excess Helmholtz free energy (left panels) and logarithm of the ionic activity coefficient (right panels) against the ionic packing fraction at various coupling constants indicated in the legends. Solid curves: SCDH predictions from Eqs.~(\ref{eq33}) and~(\ref{eq36}). Dotted curves: CCDH results from Eqs.~(\ref{eq34}) and~(\ref{eq37}). Symbols: MC simulation data from (a)-(e) Table III and (f)-(j) Table IV of Ref.~\cite{NetzMC}.}
\label{fig2}
\end{figure}

Fig.~\ref{fig1} illustrates the SCDH predictions and the MC data for the similar- and opposite-charge pair distributions in monovalent (left panel) and divalent liquids (right panel). One sees that despite the large molar concentration $n_i=1.96$ M, the SCDH approach can predict these radial distributions with quantitative precision. One also notes that in the specific case of divalent liquids, the similar- and opposite-charge distribution curves are respectively characterized by a  peak and a well located in the region $1.7\lesssim r/a\lesssim3.0$. Our formalism can accurately reproduce as well these apparent coion attraction and counterion repulsion effects originating from the onset of ion clusters composed of alternating charge layers.

\subsection{Free energy}

Figs.~\ref{fig2} (a)-(e) display the excess free energy against the ionic packing fraction at various coupling strengths indicated in the legends. The non-uniform density dependence illustrated by the plots is a manifestation of the competition between the opposite-charge attraction that dominates the system at low densities and leads to the drop of the free energy upon salt increment, and the repulsive HC interactions taking over electrostatics and leading to the rise of the free energy at large densities. From the top to the bottom plot, the increase of the coupling strength amplifies precisely the weight of the attractive electrostatic contribution and reduces the free energy of the electrolyte at all densities.

In Figs.~\ref{fig2} (a)-(e), the comparison of the theoretical curves with the MC simulation data shows that the SCDH formalism can accurately reproduce the corresponding non-uniform trend of the free energy from the WC ($\Gamma=0.5$) to the intermediate coupling regime ($\Gamma=6.0$) over the whole density range $\eta\lesssim0.15$. However, the CCDH prediction~(\ref{eq34}) remains accurate over this range only at low coupling strengths $\Gamma\lesssim1.5$; at intermediate couplings $\Gamma\gtrsim1.5$, the accuracy of the CCDH formalism is limited to the strict dilute regime of $\eta\lesssim0.01$.

\subsection{Entropy}

\begin{figure}
\includegraphics[width=1\linewidth]{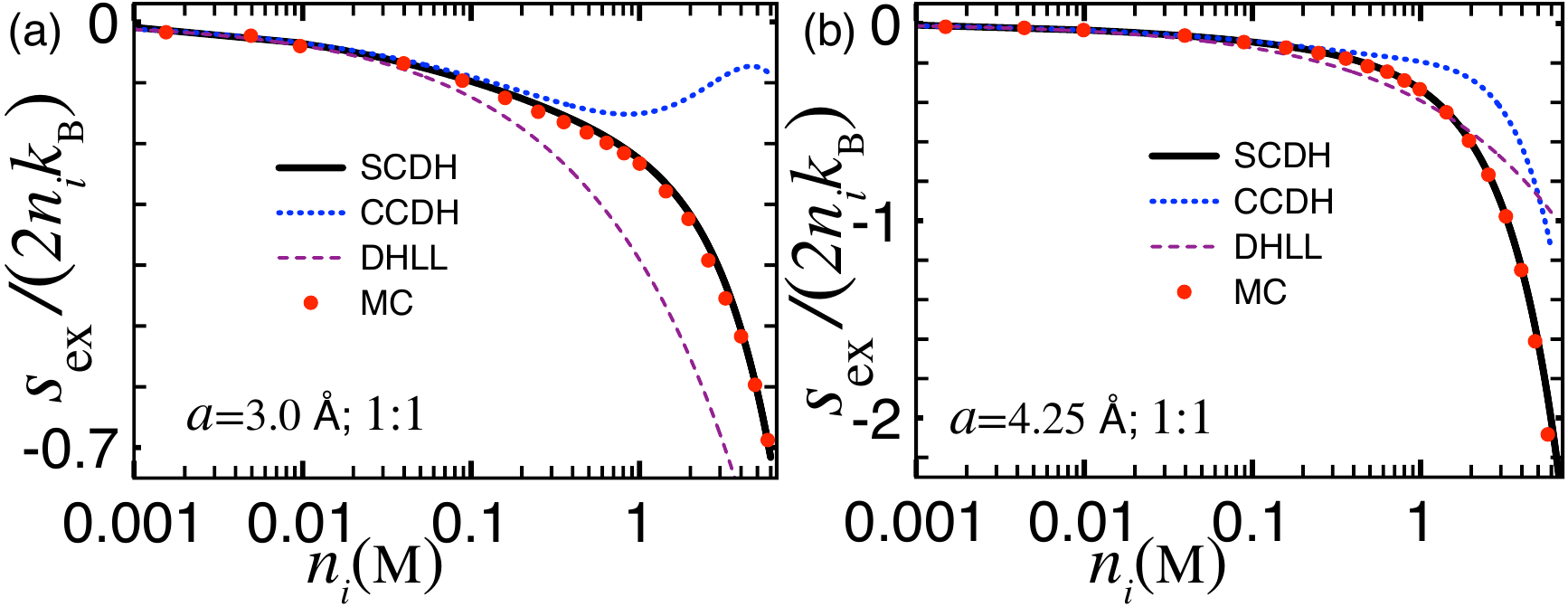}
\caption{(Color online)  Excess entropy per particle in monovalent solutions against salt concentration at temperature $T=298.15$ K, dielectric constant $\e\s=78.5$, and HC size (a) $a=3.0$ {\AA} and (b) $a=4.25$ {\AA}. Symbols: MC simulation data from Figure 2 of Ref.~\cite{Lamperski2025}. Solid curves: SCDH prediction from Eq.~(\ref{eq35}). Dotted curves: CCDH result~(\ref{eq35II}). Dashed curves: DHLL~(\ref{eq35III}).}
\label{fig3}   
\end{figure}

In order to compute the entropy $S$, we recast first the thermodynamic identity $F=U-TS$ relating the Helmholtz free energy $F$ and the internal energy $U$ in terms of the excess thermodynamic variables as $S_{\rm ex}/k_{\rm B}=\beta(U_{\rm ex}-F_{\rm ex})$. Dividing the latter equality by the volume $V$, the density of excess entropy $s_{\rm ex}=S_{\rm ex}/V$ per particle can be expressed in terms of the thermodynamic functions~(\ref{eq29}) and~(\ref{eq33}) as
\be
\label{eq35}
\frac{s_{\rm ex}}{2n_ik_{\rm B}}=\frac{\beta u_{\rm ex}}{2n_i}-\frac{\beta f_{\rm ex}}{2n_i}.
\ee

In Appendix~\ref{loop}, we derive as well a closed-form expression for entropy. Namely, by inserting into the r.h.s. of Eq.~(\ref{eq35}) the loop-expanded internal and free energies, one obtains the CCDH limit of the entropy in the form
\bea
\label{eq35II}
\frac{s^{(0)}_{\rm ex}}{2n_ik_{\rm B}}&=&-\frac{\bkap^2}{6\Gamma}-\frac{5\bkap^4}{576\Gamma^2}\\
&&+\frac{1}{2304}\left\{16-96\left[\mathrm{Ei}(-2\bkap)-\mathrm{Ei}(-\bkap)-\ln(2)\right]\right.\nonumber\\
&&\hspace{1.3cm}-8(15\bkap^3-17\bkap^2+20\bkap+32)e^{-\bkap}\nonumber\\
&&\hspace{1.3cm}+(20\bkap^3-18\bkap^2+294\bkap+243)e^{-2\bkap}\nonumber\\
&&\left.\hspace{1.3cm}-3(4\bkap+1)e^{-4\bkap}\right\}\nonumber\\
&&+\frac{\Gamma}{6912}\left\{2125+72(6\bkap^3-2\bkap^2-83\bkap-54)e^{-\bkap}\right.\nonumber\\
&&\hspace{1.3cm}-72(2\bkap^3-27\bkap^2-70\bkap-27)e^{-2\bkap}\nonumber\\
&&\hspace{1.3cm}+8(108\bkap^2-105\bkap-26)e^{-3\bkap}\nonumber\\
&&\left.\hspace{1.3cm}+9(-16\bkap^2+12\bkap+3)e^{-4\bkap}\right\}.\nonumber
\eea
Finally, in the vanishing HC size limit ($d\to0$),  the DHLL for excess entropy directly follows from Eq.~(\ref{eq35II}) as
\be
\label{eq35III}
\frac{s^{({\rm DH})}_{\rm ex}}{2n_ik_{\rm B}}=-\frac{q_i^2}{6}\kappa\ell_{\rm B}.
\ee

In Fig.~\ref{fig3}, the theoretical predictions~(\ref{eq35})-(\ref{eq35III}) are compared with the recent MC data of Ref.~\cite{Lamperski2025} at two different HC diameters. First, one sees that the DHLL law~(\ref{eq35III}) agrees with the MC result in the strictly submolar regime $n_i\lesssim0.1$ M. Then, owing to the consideration of the finite HC size, the accuracy of the CCDH result~(\ref{eq35II}) extends up to the larger concentration of $n_i\sim0.5$ M. Finally, for both ion size values, the SCDH formalism can reproduce with good quantitative precision the MC simulation result up to several molar concentrations.

\subsection{Ionic activity}

Finally, we calculate the ionic activity coefficients. To this aim, we first note that the Gibbs free energy $G$ defined as the Legendre transform $G=F+PV$ of the Helmholtz free energy $F$ is related to the chemical potential $\mu_i$ corresponding to the Gibbs energy per particle via the identity $G=\sum_i\mu_iN_i=2N_i\mu_i$. Combining the last two relations, and using the definition of the osmotic pressure  $\phi\equiv\beta P/(2n_i)$, the chemical potential takes the form $\beta\mu_i=\beta F/(2N_i)+\phi$. Finally, removing the ideal gas contribution from the latter equality, the logarithm of the activity coefficient related to the excess chemical potential $\mu_{i,{\rm ex}}$ by the relation $\ln\gamma_{\pm}=\beta \mu_{i,{\rm ex}}$ follows as~\cite{Valleau}
\be\label{eq36}
\ln\gamma_{\pm}=\phi-1+\frac{\beta f_{\rm ex}}{2n_i}.
\ee

\begin{figure}
\includegraphics[width=1\linewidth]{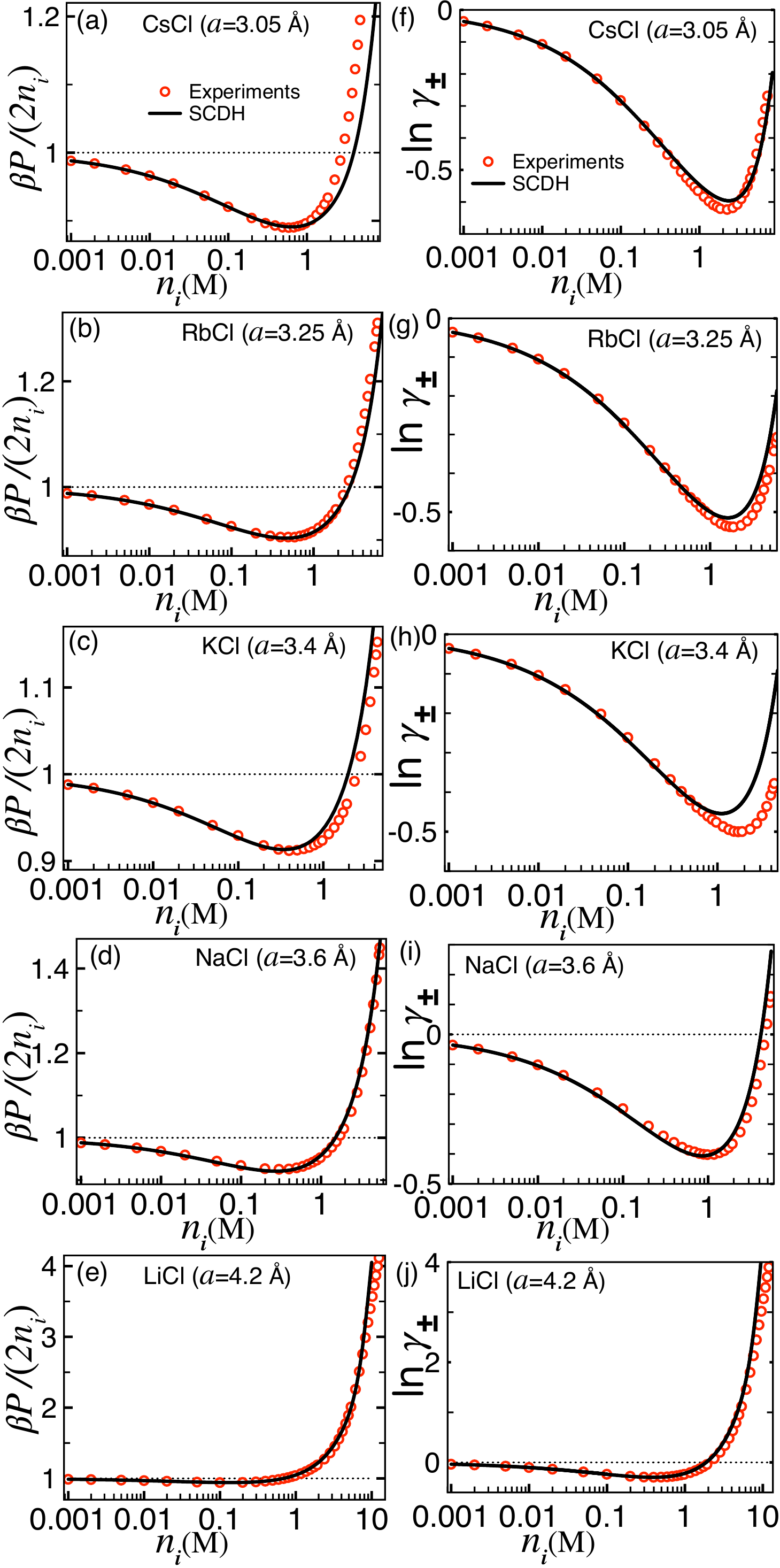}
\caption{(Color online) (a)-(e) Osmotic coefficient and (f)-(j) logarithm of the activity coefficient of various monovalent aqueous electrolytes at temperature $T=298$ K. Solid curves: SCDH results from Eqs.~(\ref{eq30}) and~(\ref{eq36}). Symbols: Experimental data from Ref.~\cite{Hamer}. The HC size values in the legends have been adjusted to obtain the best fit to the submolar concentration branch of the experimental osmotic coefficient data.}
\label{fig4}
\end{figure}

The identity~(\ref{eq36}) enables now the SCDH-level evaluation of the ionic activity coefficient via the numerical calculation of the osmotic coefficient~(\ref{eq30}) and the free energy~(\ref{eq33}). In the CCDH limit $\sigma\to0$, Eq.~(\ref{eq36}) provides as well an analytical approximation to the activity coefficient. In Appendix~\ref{loop},  we show that by plugging into Eq.~(\ref{eq36}) the loop-expanded forms of the osmotic coefficient and the Helmholtz free energy, one obtains
\bea
\label{eq37}
\hspace{1mm}\ln\gamma_\pm^{(0)}\hspace{-2mm}&=&\hspace{-2mm}\frac{\bkap^2}{3\Gamma}+\frac{5\bkap^4}{192\Gamma^2}\\
&&\hspace{-2mm}+\frac{1}{2304}\left\{16\left[6\;{\rm Ei}(-2\bkap)-6\;{\rm Ei}(-\bkap)-6\ln(2)-1\right]\right.\nonumber\\
&&\hspace{1.1cm}-8\left(5\bkap^3-19\bkap^2+4\bkap-8\right)\hspace{-.5mm}e^{-\bkap}\nonumber\\
&&\hspace{1.1cm}+\left(20\bkap^4-68\bkap^3+42\bkap^2+90\bkap-51\right)\hspace{-.5mm}e^{-2\bkap}\nonumber\\
&&\hspace{1.1cm}\left.+\left(3+12\bkap-24\bkap^2\right)e^{-4\bkap}\right\}\nonumber\\
&&\hspace{-2mm}+\frac{\Gamma}{6912}\left\{-2125+72\left(2\bkap^3-6\bkap^2-33\bkap+54\right)\hspace{-.5mm}e^{-\bkap}\right.\nonumber\\
&&\hspace{1.1cm}-72\left(2\bkap^4-8\bkap^3-16\bkap^2+6\bkap+27\right)\hspace{-.5mm}e^{-2\bkap}\nonumber\\
&&\hspace{1.1cm}-8\left(36\bkap^2+3\bkap-26\right)e^{-3\bkap}\nonumber\\
&&\hspace{1.1cm}\left.-9\left(32\bkap^3-56\bkap^2+12\bkap+3\right)\hspace{-.5mm}e^{-4\bkap}\right\}.\nonumber
\eea

Figs.~\ref{fig2}(f)-(j) illustrate the ionic activity predictions~(\ref{eq36})-(\ref{eq37}) together with the MC simulation data of Ref.~\cite{NetzMC}. One notes that the accuracy level of the theoretical ionic activity results is similar to that of the free energy functions in Figs.~\ref{fig2}(a)-(e). Namely, in the WC regime $\Gamma\lesssim1.5$, the precision of the loop-expanded result~(\ref{eq36}) extends up to $\eta\sim0.15$. Then, at intermediate couplings $\Gamma\gtrsim1.5$, this precision is limited to the strictly dilute regime $\eta\lesssim0.01$. Finally, over the whole packing fraction range covered by the MC data, the SCDH approach can reproduce with reasonable quantitative accuracy the non-uniform density dependence of the ionic activity up to the highest coupling strength $\Gamma=6.0$. 

We compare now the predictions of the SCDH theory with experimental ionic activity data from the literature~\cite{Hamer}. Fig.~\ref{fig4} displays the experimental osmotic coefficients (left plots) and ionic activities (right plots) of five different monovalent aqueous electrolytes (circles) together with the SCDH predictions (solid curves)~\cite{rem1}. In Figs.~\ref{fig4}(a)-(e), the osmotic coefficient~(\ref{eq30}) of the SCDH formalism has been plotted by adjusting the hydrated ion size to get the best agreement with the submolar branch of the experimental osmotic coefficient data. The corresponding comparison of the experimental osmotic coefficients and the SCDH result has been previously presented in Ref.~\cite{Buyuk2026}. In Figs.~\ref{fig4}(f)-(j), the ionic activity of the SCDH formalism has been calculated via Eq.~(\ref{eq36}) for the same fitted HC size values. 

In Fig.~\ref{fig4}, the comparison of the theoretical curves with the experimental data shows that with a single HC size value, the SCDH formalism can reproduce with equally reasonable accuracy the non-uniform salt dependence of the independently measured osmotic coefficients and the activity coefficients, as well as the variation of this salt dependence with the hydrated ion size from submolar to highly molar concentrations. It is also noteworthy that these fitted hydration radii lie within the range of the experimentally measured HC diameters for the corresponding ion species~\cite{Kiriukhin}. In future works, the direct incorporation of these experimentally extracted ionic diameters into our theory will necessitate the generalization of the SCDH formalism to include separately the individual hydration radius of the cationic and anionic species.

\section{Conclusions}

The quantitatively accurate evaluation of the activity coefficients of solvated ions is relevant for the characterization of the chemical reactions governing various biological and artificial nanoscale systems. Due to the particular form of Eq.~(\ref{eq32}) involving the infinite dilution limit of the electrolyte, the computation of the ionic activity coefficients via the Legendre transform of the free energy is difficult to handle numerically by computational thermodynamic theories~\cite{Soft2024,Buyuk2026} and simulations~\cite{NetzMC}.

In this work, this challenge has been overcome via the analytical evaluation of the low density branch of the integral~(\ref{eq32}) with the loop expansion of the CCDH approach~\cite{Buyuk2024} corresponding precisely to the low density limit of the SCDH theory~\cite{Soft2024,Buyuk2026}. By comparison with extensive MC simulation results from the literature~\cite{NetzMC,Lamperski2025}, we showed that the resulting identity~(\ref{eq33}) enables the quantitatively accurate calculation of the free energy and the related thermodynamic functions of aqueous electrolytes over a broad range of ion densities, HC sizes, and electrostatic coupling strengths. For the corresponding thermodynamic functions, we also derived ion size-augmented DH-like analytical expressions that may be useful to experimentalists.

First, the SCDH formalism has been shown to predict accurately the radial distributions of monovalent and divalent ions within the molar concentration regime. Then, we found that for two distinct HC sizes, the SCDH theory can reproduce with good quantitative precision the excess entropy of the liquid from the very dilute regime up to several molar concentrations. Moreover, it was shown that the our approach is able to predict as well the free energy and the ionic activities within the entire packing fraction range $\eta\lesssim0.15$ and up to the intermediate coupling strength $\Gamma\approx6.0$. We also compared the predictions of the present formalism with experimental measurements. We showed that with a single fitted HC size value, the theory can accurately reproduce the non-uniform salt dependence of the separately measured osmotic coefficients and the ionic activities of monovalent electrolytes from submolar to molar concentrations.

The close inspection of the activity coefficient plots in Figs.~\ref{fig2}(f)-(j) allows to identify the validity limit of the formalism. First, one notes that at the packing fraction value $\eta=0.15$, the SCDH curve slightly underestimates the MC data. The corresponding deviation is a consequence of the first-order cumulant-level approximation employed for the incorporation of HC correlations by the SCDH approach. Indeed, in the same high density regime of the pressure curves reported in Fig.~\ref{fig5}, one can notice a similar departure of the SCDH prediction from the simulation result. One also notes that at the highest electrostatic coupling $\Gamma=6.0$ considered in Fig.~\ref{fig2}(j), the MC prediction is marginally underestimated by the SCDH theory at all packing fraction values covered by the simulation data. This indicates that the corresponding value is the characteristic coupling strength beyond which the variational treatment of electrostatic correlations at the basis of the SCDH approach breaks down.

In future works, the accuracy of the SCDH approach can be improved by introducing more general splitting schemes, and by extending the cumulant approximation underlying our approach to higher orders. In order to account for the ion specificity beyond the primitive HC ion model, the inner charge structure~\cite{Wang} of the individual ions, their polarizability, and the non-local dielectric response of solvent can be included as well into the SCDH theory~\cite{Buyuk2013}. Generalizations of our formalism along these lines will be presented in upcoming works.

\smallskip
{\bf Acknowledgement.} I would like to thank the University of Lille and UGSF for their hospitality during a sabbatical in 2026-2027 where some parts of this work has been accomplished.

\smallskip
\appendix

\section{CCDH Limit $\sigma\to0$ of the SCDH Theory}
\label{loop}

We derive here closed-form expressions for the thermodynamic functions. To this aim,  we consider the CCDH limit of the SCDH theory corresponding to the assumption of a vanishing splitting length ($\sigma\to0$), and expand the total correlation function~(\ref{eq26}) in terms of the coupling parameter $\Gamma$. First, we note that in the corresponding limit, the short-range potential component~(\ref{eq6II}) vanishes, i.e. $v\s(u)\to0$, and the long-range potential~(\ref{eq10}) and its FT~(\ref{eq15II}) simplify to their DH limits, i.e.
\bea
\label{a1}
&&\lim_{\sigma\to0}\bar{G}_{{\rm l},0}=\frac{\Gamma}{q_i^2}\bar{g}(q);\;\;\bar{g}(q)\equiv\frac{4\pi}{\bkap^2+q^2};\\
\label{a2}
&&\lim_{\sigma\to0}G_{{\rm l},0}=\frac{\Gamma}{q_i^2}g(u);\;\;g(u)\equiv\frac{e^{-\bkap u}}{u},
\eea
where we used the dimensionless variables and parameters introduced in Sec.~\ref{adim}. Hence, expanding the Mayer function~(\ref{eq9}) at the order $O\left(\Gamma^3\right)$, one obtains
\be
\label{a3}
h_{+\pm}(u>1)\approx\mp\Gamma g(u)+\frac{\Gamma^2}{2}g^2(u)\mp\frac{\Gamma^3}{6}g^3(u),
\ee
and its FT~(\ref{eq15}) becomes
\be
\label{a4}
\bar{h}_{+\pm}\approx\bar{g}_0(q)\mp\Gamma\bar{g}_1(q)+\frac{\Gamma^2}{2}\bar{g}_2(q)\mp\frac{\Gamma^3}{6}\bar{g}_3(q),
\ee
where we introduced the functions
\bea
\label{a5}
\bar{g}_0(q)&=&-\frac{4\pi}{q^3}\left(\sin q-q\cos q\right);\\
\label{a6}
\bar{g}_{n>0}(q)&=&\frac{4\pi}{q}\int_1^\infty\mathrm{d}u'u'\sin(qu')g^n(u').
\eea
Substituting the expansions~(\ref{a1}) and~(\ref{a4}) into Eqs.~(\ref{eq14}) and~(\ref{eq16}), the auxiliary function bringing first order cumulant corrections to the total correlation function becomes
\begin{widetext}
\be
\label{a7}
T_{+\pm}(u)=2\bn\int_0^\infty\frac{\mathrm{d}qq^2}{2\pi^2}\frac{sin(qu)}{qu}\left\{\bar {g}_0^2(q)
+\Gamma^2\left[\pm \bar {g}_1^2(q)+\bar {g}_0(q)\bar {g}_2(q)\mp\bar{g}^2(q)\right]\pm2\bn\Gamma^3\bar{g}^2(q)\left[\bar{g}(k)-\bar{g}_1(k)\right]\right\}+O\left(\Gamma^4\right).
\ee
Carrying out the multiple integrals in Eq.~(\ref{a7}), and accounting for the relation $2\bn\Gamma^2/\bkap^2=\Gamma/(4\pi)$, one obtains 
\be
\label{a8}
T_{+\pm}(u)\approx2\bn\gamma(u)+\frac{\Gamma}{4\pi}\left[J_0(u)\pm J_1(u)\right],
\ee
where we defined the auxiliary functions
\bea
\label{a9}
\gamma(u)&=&\frac{\pi}{12}(u+4)(u-2)^2\theta(2-u)\theta(u-1);\\
J_0(u)&=&\frac{\pi}{4u}\left\{(2\bkap u-2\bkap-1)e^{-2\bkap(u+1)}+(-4\bkap u+2\bkap+1)e^{-2\bkap}\right.\nonumber\\
&&\hspace{7mm}\left.+4\bkap^2(u^2-1)\left[\mathrm{Ei}(-2\bkap u-2\bkap)-\mathrm{Ei}(-2\bkap)\right]\right\}\theta(2-u)\theta(u-1)\nonumber\\
&&+\frac{\pi}{4u}\left\{(2\bkap u-2\bkap-1)e^{-2\bkap(u+1)}+(-2\bkap u-2\bkap+1)e^{-2\bkap(u-1)}\right.\nonumber\\
&&\hspace{1cm}\left.-4\bkap^2(u^2-1)\left[\mathrm{Ei}(-2\bkap u+2\bkap)-\mathrm{Ei}(-2\bkap u-2\bkap)\right]\right\}\theta(u-2);\nonumber\\
\label{a10}
J_1(u)&=&\frac{\pi}{4u}\left\{(2\bkap u+2\bkap+3)e^{-2\bkap}+\left[-3+2\bkap(2\bkap-5)u+2\bkap(2-\bkap)\right]\right\}e^{-\bkap u}\nonumber\\
&&+\frac{2\pi}{u}\left\{(1-e^{-\bkap u})\theta(2-u)\theta(u-1)+\left[-e^{-2\bkap}+\bkap(u-2)+1\right]\theta(u-2)\right\}.
\eea
In Eq.~(\ref{a9}), we used the exponential integral function ${\rm Ei}(x)$ ~\cite{math}. Plugging now Eqs.~(\ref{a3}) and~(\ref{a8}) into Eq.~(\ref{eq26}), the loop-expanded total correlation function becomes
\bea\label{a11}
H_{+\pm}(u)&\approx&2\bn\gamma(u)+\Gamma\left\{\mp\left[1+2\bn\gamma(u)\right]\frac{e^{-\bkap u}}{u}+\frac{1}{4\pi}\left[J_0(u)\pm J_1(u)\right]\right\}\\
&&+\frac{\Gamma^2}{4\pi}\left\{2\pi\left[1+2\bn\gamma(u)\right]\frac{e^{-\bkap u}}{u}\mp J_0(u)-J_1(u)\right\}\frac{e^{-\bkap u}}{u}.
\eea
Finally, inserting the loop-expanded correlation functions in Eq.~(\ref{a11}) into Eqs.~(\ref{eq29})-(\ref{eq30}), and evaluating the spatial integrals, the internal energy and osmotic pressure follow in closed-forms
\bea
\label{a12}
\frac{\beta u^{(0)}_{\rm ex}}{2n_i}&=&\frac{1}{96}\left\{6(1+2\bkap)e^{-\bkap}-(6+6\bkap-9\bkap^2+5\bkap^3)\right\}e^{-\bkap}\\
&&+\frac{\Gamma\bkap}{32}\left\{2\bkap^2-2\bkap-29+8(\bkap+2)e^{-\bkap}+(4\bkap-3)e^{-2\bkap}\right\}e^{-\bkap};\nonumber\\
\label{a13}
\phi^{(0)}&=&1+\frac{\bkap^2}{6\Gamma}+\frac{5\bkap^4}{288\Gamma^2}\\
&&+\frac{1}{576}\left\{-6\bkap^2e^{-4\bkap}+(5\bkap^4-12\bkap^3+6\bkap^2+24\bkap+12)e^{-2\bkap}-2(5\bkap^3-9\bkap^2+6\bkap+6)e^{-\bkap}\right\}\nonumber\\
&&+\frac{\Gamma\bkap}{96}\left\{(5-4\bkap)\bkap e^{-4\bkap}-(4\bkap+3)\bkap e^{-3\bkap}+(-2\bkap^3+6\bkap^2+19\bkap+16)e^{-2\bkap}+(2\bkap^2-2\bkap-29)e^{-\bkap}\right\},\nonumber
\eea
where the superscript of the symbols $u^{(0)}_{\rm ex}$ and $\phi^{(0)}$ designate the CCDH limit.

Next, we introduce in Eq.~(\ref{eq32}) the change of variable $\bn\to\kappa$ and express the Helmholtz free energy in the form
\be
\label{a14}
\frac{\beta f_{\rm ex}}{2n_i}=2\int_0^{\kappa}\mathrm{d}\kappa'\frac{\phi\left(\kappa';\Gamma\right)-1}{\kappa'}.
\ee
Evaluating the integral in Eq.~(\ref{a14}) with the loop-expanded pressure~(\ref{a13}), the Helmholtz free energy follows as
\bea
\label{a15}
\frac{\beta f^{(0)}_{\rm ex}}{2n_i}&=&\frac{\bkap^2}{6\Gamma}+\frac{5\bkap^4}{576\Gamma^2}\\
&&+\frac{1}{2304}\left\{96\left[\mathrm{Ei}(-2\bkap)-\mathrm{Ei}(-\bkap)-\ln(2)\right]-16+16\left(5\bkap^2+\bkap+7\right)e^{-\bkap}\right.\nonumber\\
&&\hspace{1.3cm}\left.-\left(20\bkap^3-18\bkap^2+6\bkap+99\right)e^{-2\bkap}+3(4\bkap+1)e^{-4\bkap}\right\}\nonumber\\
&&+\frac{\Gamma}{6912}\left\{-2125-144(2\bkap^2+2\bkap-27)e^{-\bkap}+72(2\bkap^3-3\bkap^2-22\bkap-27)e^{-2\bkap}\right.\nonumber\\
&&\hspace{1.3cm}\left.+16(12\bkap+13)e^{-3\bkap}+9(16\bkap^2-12\bkap-3)e^{-4\bkap}\right\}.\nonumber
\eea
Subtracting now the internal energy~(\ref{a12}) and the free energy~(\ref{a15}), one gets from Eq.~(\ref{eq33}) the loop-expanded excess entropy as follows,
\bea
\label{a16}
\frac{s^{(0)}_{\rm ex}}{2n_ik_{\rm B}}&=&-\frac{\bkap^2}{6\Gamma}-\frac{5\bkap^4}{576\Gamma^2}\\
&&+\frac{1}{2304}\left\{16-96\left[\mathrm{Ei}(-2\bkap)-\mathrm{Ei}(-\bkap)-\ln(2)\right]-8(15\bkap^3-17\bkap^2+20\bkap+32)e^{-\bkap}\right.\nonumber\\
&&\left.\hspace{1.3cm}+(20\bkap^3-18\bkap^2+294\bkap+243)e^{-2\bkap}-3(4\bkap+1)e^{-4\bkap}\right\}\nonumber\\
&&+\frac{\Gamma}{6912}\left\{2125+72(6\bkap^3-2\bkap^2-83\bkap-54)e^{-\bkap}-72(2\bkap^3-27\bkap^2-70\bkap-27)e^{-2\bkap}\right.\nonumber\\
&&\left.\hspace{1.3cm}+8(108\bkap^2-105\bkap-26)e^{-3\bkap}+9(-16\bkap^2+12\bkap+3)e^{-4\bkap}\right\}.\nonumber
\eea
Finally, inserting the osmotic coefficient~(\ref{a13}) and the Helmholtz free energy~(\ref{a15}) into Eq.~(\ref{eq34}), the loop-expansion approximation for the logarithm of the ionic activity follows as
\bea
\label{a17}
\ln\gamma_\pm^{(0)}&=&\frac{\bkap^2}{3\Gamma}+\frac{5\bkap^4}{192\Gamma^2}\\
&&+\frac{1}{2304}\left\{16\left[6\;{\rm Ei}(-2\bkap)-6\;{\rm Ei}(-\bkap)-6\ln(2)-1\right]-8\left(5\bkap^3-19\bkap^2+4\bkap-8\right)e^{-\bkap}\right.\nonumber\\
&&\hspace{1.3cm}\left.+\left(20\bkap^4-68\bkap^3+42\bkap^2+90\bkap-51\right)e^{-2\bkap}+\left(3+12\bkap-24\bkap^2\right)e^{-4\bkap}\right\}\nonumber\\
&&+\frac{\Gamma}{6912}\left\{-2125+72\left(2\bkap^3-6\bkap^2-33\bkap+54\right)e^{-\bkap}-72\left(2\bkap^4-8\bkap^3-16\bkap^2+6\bkap+27\right)e^{-2\bkap}\right.\nonumber\\
&&\hspace{1.3cm}\left.-8\left(36\bkap^2+3\bkap-26\right)e^{-3\bkap}-9\left(32\bkap^3-56\bkap^2+12\bkap+3\right)e^{-4\bkap}\right\}.\nonumber
\eea

\end{widetext}

\section{CCDH, SCDH, and MC predictions for osmotic coefficients}
\label{comp}

\begin{figure}
\includegraphics[width=1\linewidth]{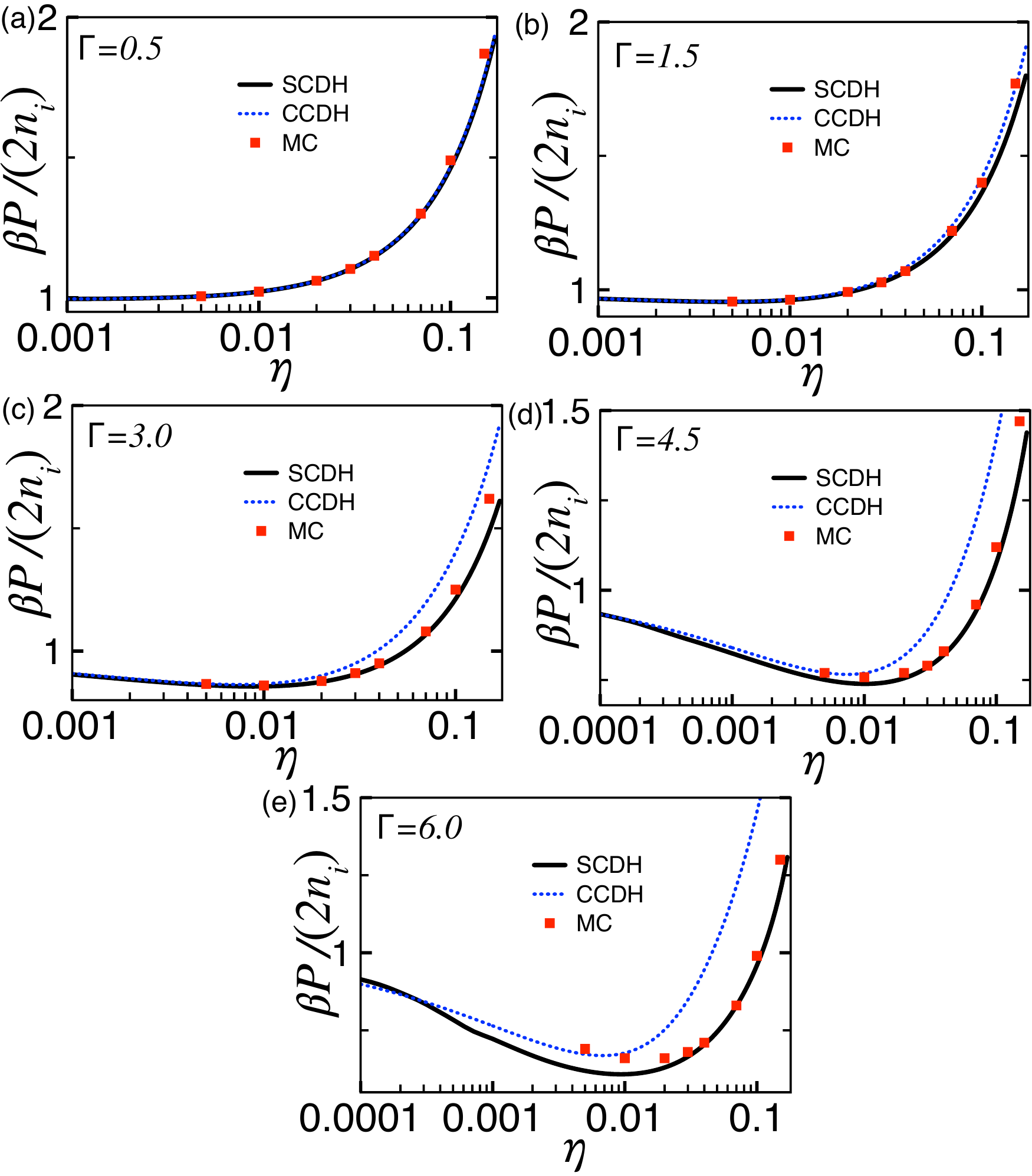}
\caption{(Color online)  Osmotic coefficient against ionic packing fraction at various coupling constants indicated in the legends. Solid curves: SCDH prediction from Eq.~(\ref{eq30}). Dotted curves: CCDH result~(\ref{a13}). Symbols: MC simulation data from Table II of Ref.~\cite{NetzMC}.}
\label{fig5}
\end{figure}

In this appendix, we report Fig.~\ref{fig5} comparing the osmotic coefficient of the CCDH approach~(\ref{a13}) (dotted curves) and the SCDH formalism~(\ref{eq30}) (solid curves) with the MC simulation data of Ref.~\cite{NetzMC} (symbols) against the ionic packing fraction $\eta$ at various coupling strengths $\Gamma$ indicated in the legends. Based on this comparison that has been previously presented in our earlier work of Ref.~\cite{Buyuk2026}, in Sec.~\ref{hel} of the main text, the packing fraction value $\eta=10^{-3}$ is chosen as the upper packing fraction below which the osmotic coefficient of the SCDH formalism in Eq.~(\ref{eq32}) is approximated by the CCDH result~(\ref{a13}).

\end{document}